\PassOptionsToPackage{english}{babel}

\documentclass[aps,pra,reprint,superscriptaddress,letterpaper,twocolumn,longbibliography,floatfix]{revtex4-1}
\usepackage{graphicx}
\usepackage{amssymb}
\usepackage[english,ngerman]{babel}

\usepackage{setspace}
\usepackage[usenames, dvipsnames, x11names]{xcolor}

\usepackage[colorlinks]{hyperref}
\hypersetup{
 colorlinks=true,
 citecolor=red,
 linkcolor=blue,
 urlcolor=cyan}

\usepackage[english]{babel}
\usepackage[utf8]{inputenc}
\usepackage{blindtext}
\usepackage{amsmath}
\usepackage[T1]{fontenc}
\usepackage{dsfont}
\usepackage{esint}
\usepackage{mathtools}

\begin{document}

\selectlanguage{english}
\title{Unconditional Fock state generation using arbitrarily weak photonic nonlinearities}
\date{\today}
\author{Andrew Lingenfelter$^{1,2}$, David Roberts$^{1,2}$, A. A. Clerk}
\affiliation{Pritzker School of Molecular Engineering, University of Chicago, Chicago, IL, USA \\
$^2$Department of Physics, University of Chicago, Chicago, IL, USA}

\begin{abstract}
We present a new mechanism that harnesses extremely weak Kerr-type nonlinearities in a single driven cavity to deterministically generate single photon Fock states, and more general photon-blockaded states.  Our method is effective even for nonlinearities that are orders-of-magnitude smaller than photonic loss.  It is also completely distinct from so-called unconventional photon blockade mechanisms, as the generated states are non-Gaussian, exhibit a sharp cut-off in their photon number distribution, and can be arbitrarily close to a single-photon Fock state. Our ideas require only standard linear and parametric drives, and are hence compatible with a variety of different photonic platforms.
\end{abstract}

\maketitle


\section{Introduction}
\label{sec:introduction}

Single-photon Fock states are a fundamental resource needed in a myriad of quantum information protocols and technologies.  There is as a result 
enormous interest in resource-friendly methods for their production \cite{EiasmanReview}.  A generic, well-studied mechanism is photon blockade \cite{imamoglu_strongly_1997}:  apply a monochromatic drive to a nonlinear photonic cavity, such that the drive is only resonant for the vacuum to one photon transition, but not for higher transitions.  While conceptually simple, this conventional photon blockade (CPB) mechanism requires the single-photon nonlinearity to be much larger than the loss rate.
This regime can be achieved in highly nonlinear cavities
incorporating single atoms \cite{birnbaum_photon_2005}, quantum dots \cite{faraon_coherent_2008} or superconducting qubits \cite{Wallraff2011,Fink2017}.  Unfortunately, this standard type of photon blockade is completely out of reach in more conventional systems that exhibit only weak nonlinearities (e.g.~optical micro or nanoresonators fabricated using materials with intrinsic $\chi^{(3)}$ nonlinearities).   

 \begin{figure}[t]
     \centering
    \includegraphics[width=0.9\columnwidth]{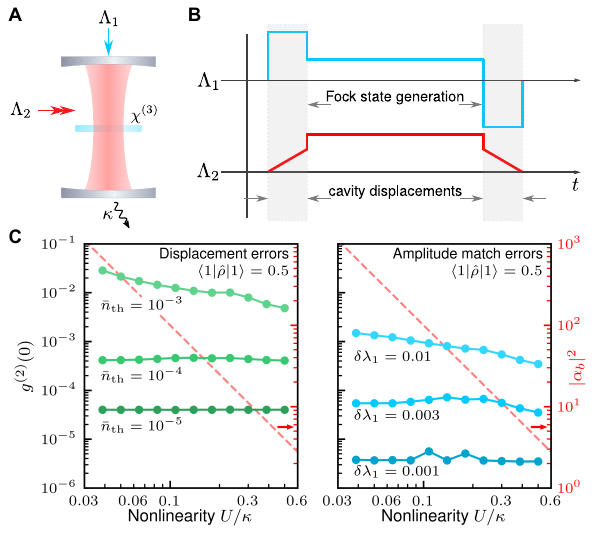}
     \caption{\textbf{Fock states with ultra-weak nonlinearities.} (a) Basic system:  a nonlinear cavity is subject to both one- and two-photon drives $\Lambda_1,\Lambda_2$.  (b) Time dependence of drive amplitudes for the protocol (see  ``Generating single photon states in the lab frame'' section).  The key idea is to realize an effective  {\it nonlinear} one photon drive in a displaced frame.  
    (c) Numerical simulations of performance including imperfections.  Parameters are chosen such that  
    the effective nonlinear drive amplitude  $\tilde{\Lambda}_3=2\kappa$, and the final state has $\langle 1|\hat{\rho}|1\rangle = 0.5$. 
    Left: $g^{(2)}(0)$ of the prepared state including errors in the initial/final displacement operations; these are modelled as added thermal noise ($\bar{n}_{\rm th}$ quanta).
    Note with added thermal noise, $g^{(2)}(0)$ must be greater than $\bar{n}_{\rm th}$.
    Right: final $g^{(2)}(0)$ with imperfect drive-amplitude matching $\delta\lambda_1\neq 0$ (cf.~Eq.~\eqref{eq:dLam1Definition}). 
    Dashed red lines show intracavity photon number $|\alpha_b|^2\sim (\kappa/U)^2$ during the intermediate part of the protocol.}
     \label{fig:introduction}
 \end{figure} 

The ability to realize effects akin to photon blockade in weakly nonlinear systems would be an incredibly powerful resource.  There has thus been a flurry of theoretical activity to uncover possible such mechanisms.  Among the best known proposals is that of ``unconventional photon blockade'' (UPB), where states with arbitrarily small $g^{(2)}(0)$ correlation functions can be generated using extremely weak nonlinearities.  UPB was 
originally proposed in Ref.~\onlinecite{liew_single_2010} and subsequently analyzed in many different works
\cite{bamba_origin_2011,flayac_input-output_2013,gerace_unconventional_2014,lemonde_antibunching_2014,zhou_unconventional_2015,wang_tunable_2015,xu_phonon_2016,flayac_unconventional_2017,sarma_unconventional_2018,hou_interfering_2019}.  It has also been realized experimentally in a circuit QED platform  \cite{vaneph_observation_2018},  and in a quantum dot plus cavity setup \cite{snijders_observation_2018}.  Unfortunately, UPB is only capable of generating Gaussian states that have positive-definite Wigner functions, and that do not exhibit a true cut-off in their photon number distribution \cite{lemonde_antibunching_2014}; moreover, they only exhibit suppressed intensity fluctuations in the limit where the average photon number is vanishingly small.  These features severely limit their utility for many possible applications.  We note that an alternative approach to stabilizing intra-cavity Fock states is to use dissipation-engineering ideas (see e.g.~\cite{Sarlette2012,mogilevtsev_nonlinear_2013,holland_single-photon-resolved_2015, souquet_fock-state_2016}).  These methods are however also resource demanding, and require strong, structured nonlinearities.   

\begin{figure}[t]
     \centering
    \includegraphics[width=0.999\columnwidth]{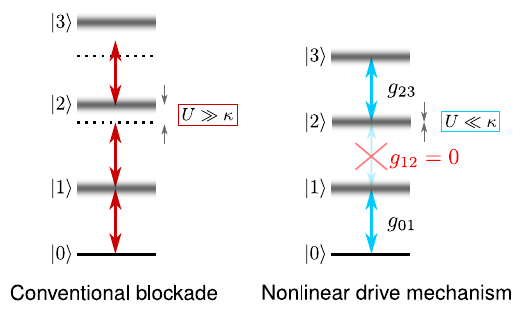}
     \caption{\textbf{Basic photon blockade mechanisms.} Left: conventional photon blockade mechanisms (CPB) rely on the nonlinearity $U$ shifting the spectrum of the system; blockade thus requires $U \gg \kappa$.  Right:  Our new approach is based on engineering a nonlinear drive that has no matrix element $g_{12}$ connecting Fock states $| 1 \rangle$ and $| 2 \rangle$.  This blockade mechanism is effective even if nonlinearity is arbitrarily weak.}
     \label{fig:cpb-vs-lam3}
 \end{figure}

In this work, we propose and analyze a new photon blockade mechanism that (unlike UPB) deterministically generates truly 
{\it non-Gaussian} blockaded states (i.e.~zero probability for more than one photon) using arbitrarily weak single-photon nonlinearities (see Fig.~\ref{fig:introduction}).  In further contrast to UPB, this can be achieved while also having the single-Fock state probability to be order 1.  Our mechanism is based on using nonlinearity to modify matrix elements of an effective cavity driving process, as opposed to introducing nonlinearity in a spectrum (as is done in CPB), see Fig.~\ref{fig:cpb-vs-lam3}.  In its simplest form, it reduces to realizing an effective single-mode Hamiltonian of the form:  
\begin{equation}
    \hat{H}_{\rm block} =\tilde{\Lambda}_3 \hat{a}^\dagger
    (\hat{a}^\dagger \hat{a} - r) + {\rm h.c.}
    \label{eq:H-blockade}
\end{equation}
where the parameter $r$ is tuned to 1.  
Here $\hat{a}$ is the cavity annihilation operator,  $\tilde{\Lambda}_3$ is the amplitude of an effective nonlinear driving process.  By construction, this Hamiltonian connects the vacuum and one photon states, but does not allow driving from $|1 \rangle$ to the $|2 \rangle$ photon state.  Crucially, as this blockade is a matrix element effect, it is effective even if cavity loss is much larger than the nonlinearity $\tilde{\Lambda}_3$.  

While the basic mechanism in Eq.~(\ref{eq:H-blockade}) is extremely simple, it describes an unusual nonlinear driving element.  At first glance, it is not at all obvious how to realize this Hamiltonian using standard $\chi^{(2)}$ or $\chi^{(3)}$ type optical nonlinearities.  Despite its exotic form, we show that it can be achieved using standard ingredients:  a standard Kerr-type nonlinearity (strength $U$), along with standard single-photon and two photon (i.e. parametric) drives.  Crucially, the mechanism is effective even if the Kerr nonlinearity strength $U$ is much much weaker than the cavity loss rate $\kappa$.  We also discuss how our scheme can be realized using three-wave mixing type (i.e.~$\chi^{(2)}$) nonlinearities.  

In what follows, we analyze in detail the physics of our basic mechanism and how it could be harnessed for a time-dependent protocol that generates propagating Fock states in a variety of realistic weakly-nonlinear optical setups.  We also discuss extensions of our basic idea, where the same underlying mechanism can be used to generate more complex blockaded states and even multi-mode non-Gaussian entangled states (see App.~\ref{app:many-body-blockade}).
Note that the infinite-time, steady-state properties of a damped cavity subject to the driving in Eq.~(\ref{eq:H-blockade}) (in a displaced frame) were studied in Ref.~\onlinecite{roberts_driven-dissipative_2020}.  While this steady state could be tuned to realize a partial blockade effect, the effect was extremely limited.  The steady state never exhibited Wigner-function negativity, and moreover was exponentially fragile to imperfections (i.e.~a small deviation of the parameter $r$ from an integer value completely destroyed the partial blockade).  The utility of this effect was thus marginal.  In contrast, our work here explores the 
{\it finite time} dynamics of systems with this kind of nonlinear driving.  We show that, surprisingly, our model exhibits metastability and two distinct slow relaxation timescales.  The intermediate-time physics can thus be extremely different from the ultimate steady state.  In particular, this regime enables the near-perfect generation of Fock states (including states with highly negative Wigner functions), in a way that is robust against imperfections.  We also stress that Ref.~\onlinecite{roberts_driven-dissipative_2020} did not discuss or analyze a concrete implementation of Eq.~(\ref{eq:H-blockade}) in a generic driven Kerr cavity system, nor did it analyze an explicit time-dependent Fock-state generation protocol; it also did not identify let alone describe quantitatively the surprising long-lived metastability of this system. These are all crucial and new features of our work.

\section{Results}

\noindent {\bf Basic mechanism and realization in a driven, weakly-nonlinear cavity.}
Despite wanting to realize a somewhat exotic nonlinear drive (cf.~Eq.~\eqref{eq:H-blockade}), we will consider a physical system that is both conventional and ubiquitous.  It consists of a single mode of a bosonic resonator (frequency $\omega_c$, lowering operator $\hat{a}$) having a weak self-Kerr nonlinearity $U$, which is subject to both one- and two-photon drives with amplitudes $\Lambda_1$ and 
$\Lambda_2$ respectively and commensurate drive frequencies $2\omega_1 = \omega_2$.
Starting from the lab-frame Hamiltonian, moving to the rotating frame set by $\omega_1$, and making a standard rotating wave approximation (RWA), we find (see ``Methods'')
\begin{equation}
    \hat{H}_{\rm RWA} = U\hat{a}^\dagger\hat{a}^\dagger\hat{a}\hat{a} + \Delta \hat{a}^\dagger\hat{a} + (\Lambda_{1}\hat{a}^\dagger + \Lambda_{2}\hat{a}^\dagger\hat{a}^\dagger + {\rm h.c.}).
    \label{eq:rwa-H}
\end{equation}
Here $\Delta = \omega_c - \omega_1$ is the detuning of the drives from cavity resonance.
We stress that the two-photon drive $\Lambda_2$ can be realized in many different ways.  For example, one could use a weak nonlinear coupling to a strongly-pumped auxiliary mode, or just simply apply two additional (linear) drive tones to the main cavity mode (see, e.g.~\cite{kamal_signal--pump_2009}).  Our results below do not depend on the specific method of implementation. 

From a quantum optics perspective, our driven cavity mode seems innocuous:  it has an extremely weak Kerr nonlinearity, and simple quadratic driving terms (which on their own would only generate simple Gaussian states).  To obtain something more interesting, our general approach is to use linear driving (i.e.~a displacement in phase space) to effectively enhance the effects of $U$. 
Such linear displacements are often used to enhance the properties of weakly nonlinear systems by yielding tuneable linear dynamics (e.g.~parametric amplifiers realized by driving weakly nonlinear cavities, or tuneable sideband interactions in quantum optomechanics \cite{AspelmeyerRMP2014}).  Such linear dynamics does not allow for the generation of non-classical, non-Gaussian states.  Here, we how a displacement can be used to generate an effective nonlinear cavity drive with a strength $\gg U$.  We note that linear driving has also been used in circuit QED experiments to generate a tuneable longitudinal coupling between a qubit and a cavity
\cite{eddins_stroboscopic_2018,touzard_gated_2019,campagne-ibarcq_quantum_2020}.  The interaction in those works is a single-photon cavity drive whose {\it phase} is controlled by an auxiliary qubit.  This is distinct from the kind of interaction we realize, namely a single-photon cavity drive whose {\it magnitude} is controlled by the photon number of the cavity itself, as opposed to that of a highly nonlinear auxiliary system.

We show that by moving to a displaced frame of the cavity, $\hat{a}\rightarrow \hat{a} + \alpha$ where $\alpha$ is an arbitrary displacement parameter, we can generate a displacement-enhanced nonlinearity that is precisely the term we seek to engineer (see ``Methods''). Upon moving to a displaced frame of the cavity, we find that the Kerr nonlinearity generates, among corrections to the other terms in $\hat{H}_{\rm RWA}$, the desired nonlinear drive $\tilde{\Lambda}_3\hat{a}^\dagger\hat{a}^\dagger\hat{a} + {\rm h.c.}$ with drive amplitude: $\tilde{\Lambda}_3 = 2 U \alpha$ (see ``Methods'',  Eq.~(\ref{eq:Lam3Tuning})).

Our goal is to realize (in our displaced frame) the ideal blockade Hamiltonian
\begin{equation}
    \hat{H}_{\rm target} = \left( \tilde{\Lambda}_3 \hat{a}^\dagger
    (\hat{a}^\dagger \hat{a} - r) + {\rm h.c.} \right) + U\hat{a}^\dagger\hat{a}^\dagger\hat{a}\hat{a}.
    \label{eq:H-blockade-target}
\end{equation}
To achieve this, we first decide on a desired strength for the nonlinear drive amplitude $\tilde{\Lambda}_3$ in  $\hat{H}_{\rm target}$, and pick the displacement parameter $\alpha$ to achieve this.  This requires:
\begin{equation}
    \alpha \rightarrow \alpha_b \equiv \frac{\tilde{\Lambda}_3}{2U}.
    \label{eq:alpha-parameter}
\end{equation}
We will typically want $\tilde{\Lambda}_3 \gtrsim \kappa$, implying that a large displacement will be needed if the nonlinearity $U$ is weak.

The last step is to pick our original drive parameters
$\Lambda_1, \Lambda_2, \Delta$ to make the remaining terms in the full displaced Hamiltonian $\hat{H}_{\alpha}$ (see Eq.~(\ref{eq:displaced-H}) in ``Methods'') match 
$\hat{H}_{\rm target}$.
This leads to the choices:
\begin{subequations}
\label{eqs:OptimalParams}
\begin{align}
    \Lambda_{1} \rightarrow \Lambda_{1,b} &\equiv \tilde{\Lambda}_3\left[ -r + \frac{|\tilde{\Lambda}_3|^2}{2U^2} + \frac{i\kappa}{4U} \right], \label{eq:Lam1-parameter} \\
    \Lambda_{2} \rightarrow \Lambda_{2,b} &\equiv -\tilde{\Lambda}_3^2 / 4U, \label{eq:Lam2-parameter} \\
    \Delta \rightarrow\Delta_b &\equiv - |\tilde{\Lambda}_3|^2 / U. \label{eq:Delta-parameter}
\end{align}
\end{subequations}

With this choice of drive parameters and displacement parameter $\alpha$, our displaced-frame Hamiltonian
$\hat{H}_\alpha$ has exactly the desired form of the target blockade-producing Hamiltonian in Eq.~\eqref{eq:H-blockade-target}.  If we pick $r$ in Eq.~\eqref{eq:Lam1-parameter} to be an integer, it follows that we can achieve blockaded dynamics in the displaced frame.  To be concrete, imagine we tune parameters to achieve $r=1$.  
If we then start the system in the vacuum of the displaced frame (i.e.~a coherent state in the lab frame), then the full system dynamics will be confined to the Fock states $n=0, n=1$ in the displaced frame, {\it regardless of how small the original value of $U$ was}.  

We have thus demonstrated how the basic physics of Eq.~\eqref{eq:H-blockade} can be realized using an arbitrarily-weak Kerr nonlinearity and standard one and two photon driving processes.  Note that the magnitude of the nonlinear driving in the displaced frame is the product of the original Kerr nonlinearity $U$ (which could be extremely small) and the displacement $\alpha$ (which at this stage, we can assume to be very large).  There is of course an important caveat about our scheme at this stage:  as described, it only yields blockaded states and Fock states {\it in the displaced frame}.  As we show in the ``Generating single photon states in the lab frame'' section below, this is not a true limitation, as we can easily harness this physics to generate true lab-frame Fock states (see also Fig.~\ref{fig:introduction}).


\noindent {\bf Blockade dynamics in the presence of loss.}
Before addressing how one converts displaced-frame blockaded states into truly blockaded states, we first investigate the dynamics of our system in the displaced frame.  We thus study displaced-frame master equation
\begin{equation}
    \frac{d}{dt}\hat{\rho} = -i[\hat{H}_{\rm target},\hat{\rho}] + \kappa\mathcal{D}[\hat{a}]\hat{\rho}
    \label{eq:blockade-qme2}
\end{equation}
where $\hat{H}_{\rm target}$ is given by Eq.~\eqref{eq:H-blockade-target}.  We will consider the dynamics when the parameter $r$ is close to, but not identical, to its ideal value for an $n=1$ Fock state blockade, i.e. $r = 1 + \delta r$.  In practice, $\delta r$ corresponds to a failure to exactly match the one and two photon drive amplitudes in the ideal required manner, as dictated by Eqs.~(\ref{eq:Lam1-parameter}) and (\ref{eq:Lam2-parameter}).  
Our focus here will be primarily on understanding the temporal dynamics on time scales $t \lesssim 1/\kappa$, and using this to identify optimal parameters for generating Fock states.  


\noindent {\bf Dynamics for ideal drive amplitude matching.}
For perfect parameter tuning $\delta r = 0$, we have ideal blockade dynamics where the drive cannot connect the $n=1$ and $n=2$ Fock states.  Within the blockade manifold spanned by $\{|0\rangle,|1\rangle\}$, the cavity behaves like a two-level-system which is resonantly driven with Rabi frequency $\propto \tilde{\Lambda}_3$, i.e.
$\hat{H}_{\rm target} \rightarrow \tilde{\Lambda}_3 |1\rangle\langle0| + \textrm{h.c.}$. 
As there is no probability of having 2 or more photons, for this perfect tuning of $r$, the equal-time $g^{(2)}$ correlation function
(defined as $g^{(2)}(0) \equiv \langle \hat{a}^\dagger \hat{a}^\dagger \hat{a} \hat{a} \rangle / \langle \hat{a}^\dagger \hat{a}\rangle^2$) is always exactly 0.  To generate a single-photon state, we simply need to perform an effective $\pi$-pulse.  This amounts to turning on the one and two photon drives (with the ideal amplitudes given by Eqs.(\ref{eq:Lam1-parameter}) and (\ref{eq:Lam2-parameter})) for a time  $t_\pi=\pi/(2|\tilde{\Lambda}_3|)$. This allows the perfect generation of a Fock state in the limit where $t_\pi \ll 1 / \kappa$, requiring $|\tilde{\Lambda}_3|/\kappa \gg 1$.  We stress that this condition can be met {\it even if}  $U \ll \kappa$.


 \begin{figure}[t]
    \centering
    \includegraphics[width=0.99\columnwidth]{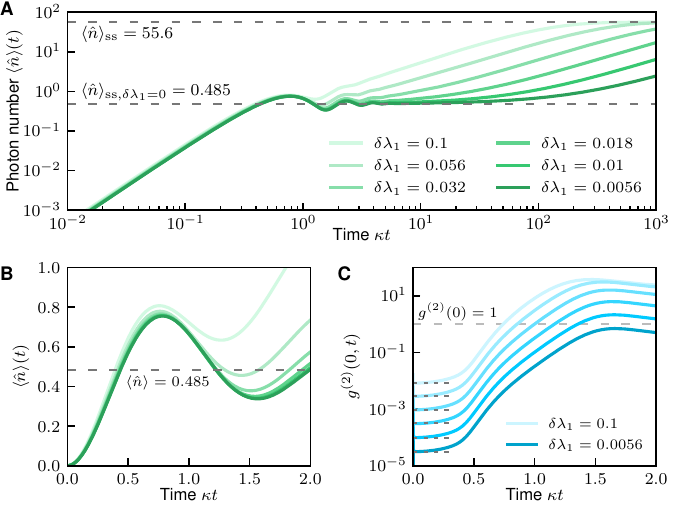}
    \caption{\textbf{Impact of mismatched drive amplitudes on blockade dynamics.} (a) Average intracavity photon number versus time (log axes) for values of the dimensionless relative amplitude mismatch $\delta\lambda_1$ 
    (cf.~Eq.~\eqref{eq:dLam1Definition}).
    One clearly sees two distinct timescales:  the desired low-amplitude blockaded state is reached on a time scale $\sim 1/\kappa$, whereas if $\delta \lambda_1 \neq 0$, there is a much slower heating to a high amplitude state, $\Gamma_{\rm esc}$ (cf.~Eq.~(\ref{eq:escape-rate})). Note that all of the $\delta\lambda_1$ shown are much larger than the ``anti-resonance'' width $\Delta r$ for these parameters (cf.~Fig.~\ref{fig:app-a}), i.e., the steady state blockade is destroyed for all $\delta \lambda_1$ shown.
    (b) Zoom in on short-time behaviour of (a), linear axes.  
    The dashed curve is the ideal $\delta \lambda_1 = 0$ steady-state average photon number.
    (c) The instantaneous intracavity correlation function $g^{(2)}(0;t)$, for various imperfection levels $\delta \lambda_1$.  Dashed lines correspond to the short-time analytic result in  Eq.~\eqref{eq:g2-three-level}. 
    For all plots we use parameters $U = 0.4\kappa$ and $\tilde{\Lambda}_3 = 2\kappa$.}
    \label{fig:blockade-breakdown}
 \end{figure} 

\noindent {\bf Impact of imperfect drive-amplitude matching.}
We now consider what is likely the dominant error mechanism for our scheme: the inability to perfectly match the drive amplitudes $\Lambda_1$ and $\Lambda_2$ as required to achieve $r=1$.  
For small mismatch $\delta r$, there is only a weak matrix element connecting $|1\rangle$ to $|2\rangle$.  As we will show, this means that we still have approximate blockade physics over a long timescale, enabling the production of non-classical blockaded states.  
The perfect single photon blockade we desire requires matching the linear and cubic driving terms in the displaced-frame Hamiltonian $\hat{H}_{\alpha}$ (cf.~Eq.~(\ref{eq:displaced-H})), i.e.~$\tilde{\Lambda}_1 = -\tilde{\Lambda}_3$ (i.e.~$r=1$).  Deviations from this amplitude-matching condition will then degrade our scheme.  We thus 
define $\delta \lambda_1$,  the dimensionless relative amplitude error in the single-photon drive amplitude, via
\begin{equation}
    \label{eq:dLam1Definition}
    \tilde{\Lambda}_1 = -\tilde{\Lambda}_3(1 +
    \delta \lambda_1). 
\end{equation}
While in general both the magnitude and phase of $\delta \lambda_1$ are important, for the small deviations we focus on here, only the magnitude matters. We take $\delta\lambda_1$ real and positive for all of the numerical simulations. 

To get some analytic insight into the impact of this imperfection, consider the most interesting regime of small imperfection $|\delta \lambda_1| \ll 1$ and large effective driving,   $|\tilde{\Lambda}_3|>\kappa$.  For short times, dissipation can be neglected, and further, the dynamics will be restricted to the states $|0\rangle$, $|1\rangle$, and $|2\rangle$ (as the leakage to higher levels is weak).  In this regime, we find that the instantaneous $g^{(2)}(0;t)$ is time-independent and given by      
\begin{equation}
    g^{(2)}(0;t) = |\delta\lambda_1|^2.
    \label{eq:g2-three-level} 
\end{equation}
This suggests that highly blockaded states are possible without requiring an incredibly precise balancing of drive amplitudes.

In Fig.~\ref{fig:blockade-breakdown}(b)-(c),
we show the results of a numerical simulation of the effects of a non-zero drive-amplitude mismatch $\delta \lambda_1$. We see that the intracavity average photon number shown in Fig.~\ref{fig:blockade-breakdown}(b) undergoes Rabi oscillations before leaving the blockaded subspace; we also see that Eq.~(\ref{eq:g2-three-level}) provides a good description of the intracavity $g^{(2)}(0)$ until a time 
$t\sim 1/|\tilde{\Lambda}_3|$, after which there is a departure from the blockaded subspace.
The net result of our simulations and analysis is that errors in amplitude matching do not prevent the generation of useful blockaded states:  for short times, the evolution produces states with small $g^{(2)}(0)$ while at the same time having appreciable non-vacuum population.  As the Figure shows, even for relative mismatches of $\delta \lambda_1 \sim 0.1$, blockaded states with $\langle \hat{a}^\dagger \hat{a} \rangle \sim 0.5$ and $g^{(2)}(0) < 0.1$ can be produced.

\noindent {\bf Slow time scales, metastability and blockaded states in the infinite-time limit.}
While for applications, the relatively robust blockade physics we obtain at short times is more than sufficient, it is also interesting to ask about the nature of the long-time steady state.  For $\delta \lambda_1 = 0$, the blockade is perfect for all times, and the steady state has no population of higher Fock states.  With imperfections, the situation is different.  
We saw above that the short-time blockade physics is relatively robust against amplitude mismatch errors.  This however is not true for the infinite-time state.
As discussed in ``Methods'', for $\delta \lambda_1 = 0$, the system has a long-lived, metastable high-photon number state that is only able to decay via quantum tunneling.  This manifests itself as an extremely slow relaxation rate (i.e.~dissipative gap):
\begin{equation}
    \gamma_{\rm slow} \sim \kappa \exp\left(-\frac{9|\tilde{\Lambda}_3|^2}{4U^2}\right)
\end{equation}
(cf.~Eq.~(\ref{eq:slow-rate}) and preceding discussion in ``Methods''). This exponentially small dissipative gap directly leads to the extreme fragility of the steady-state photon blockade to even minuscule mismatches of drive amplitude.  A simple perturbative argument suggests that the steady state blockade is lost when $|\delta \lambda_1| \simeq \gamma_{\rm slow} / \kappa $, i.e.~even when $|\delta \lambda_1| \ll 1$  (cf.~Fig.~\ref{fig:app-a}).  This fragility makes the steady-state effect essentially unattainable in experiment. 
Note that the extreme sensitivity of the steady state to relative drive amplitudes was first observed without explanation in Ref.~\cite{roberts_driven-dissipative_2020};  the qualitative and quantitative explanations of this phenomenon provided in ``Methods'' is however new to this work.

One might worry that this small dissipative gap should also have made the finite-time blockade physics presented above  highly fragile.  This is not the case:  for an imperfect system that starts from vacuum, there is a {\it distinct} metastable regime of relevance whose physics is controlled by a new timescale unrelated to $1/\gamma_{\rm slow}$.  The relevant rate $\Gamma_{\rm esc}$ now corresponds to a slow escape from the blockaded subspace. For imperfect amplitude matching ($\delta \lambda_1 \neq 0$), there is a weak coupling between blockaded and un-blockaded subspaces.  Once in the un-blockaded subspace, the system can eventually populate the weakly metastable, high-amplitude state.  While this escape destroys the blockade and results in a very large average photon number in the steady state, this corruption occurs over a very slow timescale $1/\Gamma_{\rm esc}$.  The slow heating associated with this phenomena can be seen in Fig.~\ref{fig:blockade-breakdown}(a).

The escape rate $\Gamma_{\rm esc}$ can be estimated using a Fermi's Golden Rule (FGR) argument where $\delta \lambda_1$ (the imperfection in the single photon drive amplitude) is treated as a perturbation.  This is consistent with the numerically observed behaviour that the average intracavity photon number approaches its steady-state value exponentially.  Defining $\delta\tilde{\Lambda}_1 = \tilde{\Lambda}_3 \times \delta\lambda_1$, an approximate FGR calculation yields (see Methods)
\begin{equation}
    \Gamma_{\rm esc} = c \frac{|\delta \tilde{\Lambda}_1|^2}{\kappa},
    \label{eq:escape-rate}
\end{equation}
with $c$ is a dimensionless number.  While in general it will depend on other parameters in the unperturbed Hamiltonian, for $\kappa \gg \tilde{\Lambda}_3$ we find it is constant: $c=1$. In contrast, for the regime of interest $\kappa \sim \tilde{\Lambda}_3$, a simple analytic estimate is not possible.  We do however find from numerics in this regime (i.e.~by fitting the long-time relaxation of the average photon number shown in Fig.~\ref{fig:blockade-breakdown}(a)) that $c \approx 0.25$ in this regime.  The overall form of $\Gamma_{\rm esc}$ reflects two basic facts:  the cavity can only leave the blockade subspace through the very small matrix element $\propto \delta\tilde{\Lambda}_1$, and the cavity must jump into energy eigenstates which are not localized to the Fock state $|2\rangle$ but spread out in Fock space and thus harder to jump into. The latter effect leads generically to $c < 1$.

The slow escape rate $\Gamma_{\rm esc}$ defines a time window over which the blockaded subspace is isolated from the rest of Hilbert space.  In order to prepare Fock states, one just needs this time to be long compared to inverse drive amplitudes.  In practice, this leads to the weak constraint on drive-amplitude matching $\delta \lambda_1 < 1$.  This is to be contrasted against the exponentially more demanding condition needed for blockade physics in the steady state, 
$\delta \lambda_1 < \gamma_{\rm slow} / \kappa$.
The vast difference in these conditions means that our blockade mechanism is with reach of various experimental platforms, whereas in contrast the steady-state version of the effect is completely impractical.

\noindent {\bf Photon blockade with weak drive.}
The short-time blockade physics we have considered so far requires $\tilde{\Lambda}_3 > \kappa$.  Via Eq.~(\ref{eq:alpha-parameter}), we see this is possible even if $U \ll \kappa$, as long as we use a large displacement $\alpha_b$.  While at a fundamental level such large displacements pose no problems, at a practical level they can create issues.  We will see this explicitly in the next section, where we discuss in detail how to turn the displaced-frame Fock states produced by Eq.~(\ref{eq:H-blockade-target}) to true lab-frame Fock states.  

\begin{figure}[t]
    \centering
    \includegraphics[width=0.99\columnwidth]{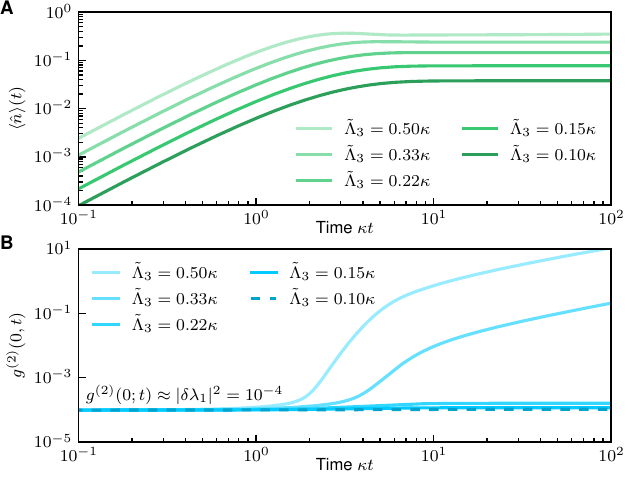}
    \caption{\textbf{Photon blockade dynamics with weak $\tilde{\Lambda}_3$ drive.}
    As discussed in the text, the resource requirements of our scheme are greatly reduced if one only tries to achieve a nonlinear drive $\tilde{\Lambda}_3 \lesssim \kappa$.
    (a) Average intracavity photon number versus time for an imperfect drive amplitude matching $\delta\lambda_1=0.01$, for different $\tilde{\Lambda}_3$.
    As expected, $\langle \hat{n}\rangle(t)$ approaches its steady state value (cf.~Eq.~\eqref{eq:damped-ss-P1}) in a time $\sim 1 / \kappa$.  Reducing  $\tilde{\Lambda}_3$ reduces this value.
    (b) Instantaneous intracavity $g^{(2)}(0;t)$ of the cavity as a function of time, with $\delta\lambda_1=0.01$.
    Even for modest drives $\tilde{\Lambda}_3 < \kappa$, a good blockade is achieved at short times.  
    For all plots $U = 0.075\kappa$.}
    \label{fig:blockade-weak-drive}
 \end{figure}

Given this possible concern, it is also interesting to ask about the dynamics of system where $|\tilde{\Lambda}_3| \ll \kappa$, a regime that could be reached with small $U$ and modest displacements $\alpha$.  Consider first the case where the drive amplitudes are perfectly matched, implying $r=1$ in Eq.~(\ref{eq:H-blockade-target}).  In this case, the system approaches the infinite-time, perfectly-blockaded steady state on a timescale $\sim 1 / \kappa$.  This state has zero probability for having more than one photon, and the single photon occupancy is
\begin{equation}
    \langle 1 | \hat{\rho}(t \rightarrow \infty) | 1 \rangle
    = \frac{4|\tilde{\Lambda}_3/\kappa|^2}{1+8|\tilde{\Lambda}_3/\kappa|^2}.
    \label{eq:damped-ss-P1}
\end{equation}
Hence, having a weak $\tilde{\Lambda}_3 / \kappa$ does not break the blockade, but just reduces the population of the one photon state.  On the bright side, in this weak drive regime, the blockade much more robust to amplitude mismatch errors. Fig.~\ref{fig:blockade-weak-drive}(a) shows the transition from the underdamped regime
$\tilde{\Lambda}_3 > \kappa/4$, where coherent oscillations are visible, to the overdamped regime where the cavity exponentially relaxes to the steady state. The robustness of the overdamped blockade is shown in Fig.~\ref{fig:blockade-weak-drive}(b) where the $g^{(2)}(0;t)$ of the overdamped blockade remains near the amplitude-mismatch-limited value $g^{(2)}(0;t) = |\delta\lambda_1|^2$ given by Eq.~\eqref{eq:g2-three-level} for long times even as the underdamped blockade experiences a large rise in $g^{(2)}(0;t)$ for times $\kappa t \sim 1$.  

\noindent {\bf Generating single photon states in the lab frame.}
Our discussion so far has established how, using a cavity mode with an extremely weak Kerr nonlinearity $U \ll \kappa$ and standard one and two photon drives, it is possible to generate truly photon-blockaded states {\it in a displaced frame}.  In the displaced frame, and for ideal matching of drive amplitudes, these states have zero population of states with two or more photons, and moreover, can have a population of the $| 1 \rangle$ Fock state that approaches one.  We also showed that this physics is robust again modest errors in matching the two drive amplitudes appropriately.  

We discussed the displacement transformation $\hat{a} \rightarrow \hat{a} + \alpha$ that led to the Hamiltonian in Eq.~(\ref{eq:displaced-H}) as a {\it passive} transformation. In order to make use of this idea to generate {\it true} Fock states, we now view the displacement as an {\it active} transformation: a short, high-amplitude one photon drive will be used to initially and rapidly displace the cavity state by an amplitude $\alpha_b$.  A similar protocol will then be used to undo this displacement at the end of the blockade protocol. In what follows, we discuss each step of this protocol in detail, including a treatment of new error mechanisms associated with imperfect displacements.  

\begin{figure}[t]
    \centering
    \includegraphics[width=0.999\columnwidth]{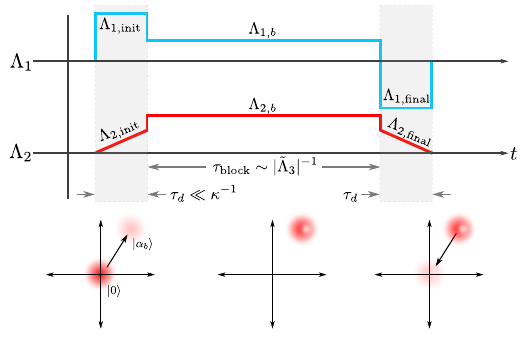}
 \caption{{\bf Fock state generation protocol timing diagram.} Upper panel:  the three steps of the generation protocol described in the ``Protocol overview'' section. The grey regions are the initial and final cavity displacements which are implemented by applying strong one photon drives ($\Lambda_1$) to the cavity for a short displacement time $\tau_d\ll\kappa^{-1}$.  Ramped two photon drives ($\Lambda_2$) are also applied to correct unwanted squeezing generated by $U$ during these displacement operations.
 The white region represents the displaced-frame Fock state generation step; here, one and two photon drive amplitudes  are tuned to their ideal values as given by Eqs.~\eqref{eq:Lam1-parameter}-\eqref{eq:Delta-parameter}.  The evolution here occurs for a duration $\tau_{\rm block}\sim|\tilde{\Lambda}_3|^{-1}$ that can be optimized, during which the cavity evolves under Eq.~(\ref{eq:blockade-qme}).  Bottom panel:  Cavity phase space diagram showing schematically the evolution of the cavity state in the lab frame.}
 \label{fig:single-photon-protocol}
 \end{figure}

\noindent {\bf Protocol overview.}
The basic idea of the full scheme is sketched  
in Fig.~\ref{fig:single-photon-protocol}.  It has three main steps: 
\begin{enumerate}
    \item \textbf{Initial displacement:} With the cavity initially in vacuum $|0\rangle$, we rapidly displace the cavity (using the one photon drive) to the coherent state $|\alpha_b\rangle$ (see Eq.~\eqref{eq:alpha-parameter}).
    \item \textbf{Fock state generation:} We next turn on the two photon drive, and set both the drive amplitudes $\Lambda_{1},\Lambda_{2}$ to their ideal values given by
    Eqs.~(\ref{eq:Lam1-parameter}) and (\ref{eq:Lam2-parameter}).  We then let the system evolve for an optimally-chosen time $\tau_{\rm block} \sim |\tilde{\Lambda}_3|^{-1}$.  
    This will prepare to good approximation a single photon blockaded state in the displaced frame.
    \item \textbf{Final displacement:} Finally, we turn off the two photon drive, and adjust the amplitude of the one photon drive $\Lambda_1$ such that rapidly displaces the cavity by an amount $-\alpha_b$.  This then shifts our displaced-frame blockaded state to lab-frame blockaded state (ideally the state $| 1 \rangle$).
\end{enumerate}
The end result of the three steps above is a blockaded, approximate single-photon state in the cavity.  To turn this into a more useful propagating single photon state, we imagine a situation where the cavity is overcoupled to a waveguide or transmission line.  In this case, one simple waits at the end of step three.  The intracavity state will then preferentially leak out into waveguide as an approximate Fock state in a propagating mode with an exponential profile.  Note that while overcoupling will increase $\kappa$, this is not overly detrimental to our protocol. As we have stressed, our protocol can be effective even if the Kerr nonlinearity $U$ is much smaller than the total loss rate $\kappa$ of the cavity.    

The initial and final displacements in our protocol are of course key aspects needed to achieve our final, lab-frame photon-blockaded state.  As discussed, these should correspond to amplitudes $\alpha_b, -\alpha_b$ respectively, where this amplitude is determined by Eq.~(\ref{eq:alpha-parameter}).  A failure to perform this ideally represents another possible experimental imperfection that would degrade from our scheme.  Even if the one photon drive used to perform these displacements can be calibrated perfectly, the weak cavity nonlinearity $U$ can cause errors during steps 1 and 3 of the protocol.  The dominant error is an unwanted parametric drive generated via $U$; this could be cancelled by also applying a compensating two photon drive $\Lambda_2 \neq 0$ during steps 1 and 3; this is depicted in
Fig.~\ref{fig:single-photon-protocol}.  In what follows, as opposed to focusing on a particular mechanism, we use a general model to characterize errors in the displacement steps (steps 1 and 3) of our general protocol.

\noindent {\bf Numerical results.}
Having outlined our full protocol, we numerically study its performance.  Step 2 is modelled exactly, by evolving our system as per the full master equation in Eq.~(\ref{eq:blockade-qme}).
The possibly imperfect displacement operations in steps 1 and 3 are modelled as a combination of a perfect displacement and the injection of thermal noise (corresponding to $\bar{n}_{\rm th}$ thermal quanta).  Formally, this corresponds to a Gaussian additive noise channel \cite{weedbrook_gaussian_2012}.  Note that this additive thermal noise rapidly degrades the blockade.  If we start with a perfect Fock state $|1 \rangle$ and add $\bar{n}_{\rm th}$ thermal quanta (via an additive Gaussian noise channel), then one can show that $g^{(2)}(0) \geq 4\bar{n}_{\rm th}$. Further details are provided in App.~\ref{app:error-modeling}, as are results for limitations arising from classical displacement and phase noise.

In addition to displacement errors, we consider drive amplitude mismatches which we discussed in the ``Impact of imperfect drive-amplitude matching'' section. The results of that analysis apply here, but as a check we perform the full Fock state generation protocol with small $\delta\lambda_1\neq 0$. The figure of merit for the Fock state generation protocol is the instantaneous second order coherence $g^{(2)}(0)$ at the end of the protocol as a function of $U/\kappa$.

Shown in Fig.~\ref{fig:introduction}(c) are numerical simulations of our full time-dependent protocol for various choices of $U / \kappa$.  In each case, parameters are chosen to produce (in the ideal case) a state where the blockaded state has $\langle 1 | \hat{\rho} | 1 \rangle = 0.5$.
The numerical results show that the blockade protocol is effective even for $U / \kappa \sim 0.03$, and moreover, is robust against  both small displacement errors and small amplitude match errors.  There is no fundamental limit against applying our protocol for even smaller values of $U$.  Numerics becomes somewhat unwieldy, given the large displacements $\alpha_b \sim \kappa / U$ that are required.   


\section{Discussion}

A key virtue of our scheme is that it is extremely generic:  there are many different kinds of systems that can realize weakly-nonlinear electromagnetic modes with one- and two-photon drives.  In the context of weakly nonlinear optical cavities, the primary experimental challenge for implementation is the large cavity displacements required, 
$\alpha_b\sim \kappa/U$.  For typical low-loss silicon micro-resonators, the intrinsic $\chi^{(3)}$ nonlinearity yields $U/\kappa\sim 10^{-8}$ \cite{vernon_spontaneous_2015}. The $\chi^{(3)}$ of silicon nitride is typically even smaller  \cite{ikeda_thermal_2008,ye_low-loss_2019,tsang_nonlinear_2008}.  While the large displacements and intra-cavity powers required in such systems to achieve $\alpha_b\sim \kappa/U$ may be possible given the pulsed nature of our scheme, a safer route would be to follow the general ideas in the ``Photon blockade with weak drive'' section.  Here, one uses displacements much smaller than $\kappa/U$, making constraints on power handling much more reasonable.  This results in a perfect blockade and states with vanishingly small $g^{(2)}(0)$.  The price to pay however is that the average photon number will also be very small.  We stress that even in this regime, the states generated have a strong advantage over the unconventional photon blockade (UPB) mechanism of Ref.~\cite{liew_single_2010}:  unlike UPB, our states are non-Gaussian and have zero population of higher Fock states.

An alternative route for implementation in optical cavities would be to utilize $\chi^{(2)}$ nonlinearities in materials with broken inversion symmetry like silicon nitride or aluminum nitride.
These nonlinearities are parametrically larger than the corresponding $\chi^{(3)}$; a recent experiment even achieved a single-photon $\chi^{(2)}$ nonlinearity that was $\sim 0.01 \kappa$ \cite{lu_toward_2020}.  We stress that while our scheme requires a Kerr-type four-wave mixing nonlinearity, this can be achieved starting with three-wave mixing $\chi^{(2)}$ processes that generates a nonlinear coupling to a detuned auxiliary mode \cite{guo_all-optical_2018}.  To second order in this coupling, one generates the desired self-Kerr interaction $U$ needed for our scheme.  Despite being second order, this can still be orders-of-magnitude larger than an intrinsic $\chi^{(3)}$ nonlinearity.

While optical cavities are one possible domain of application, they are not the only candidate.  Our ideas could also be exploited in parametrically-driven nanomechanical systems with weak intrinsic Duffing nonlinearities (see e.g.~\cite{WeigPRX2020}), as well as in microwave cavity systems.  A current trend in quantum information processing with superconducting circuits is to store and process information in high-$Q$ microwave cavities (see, e.g.~\cite{Puri2020,Grimm2020}).  In such schemes, detuned qubits are often used to induce weak nonlinearities in the principle bosonic modes.  A key limitation in these approaches is that the qubit also induces new loss mechanisms.  Our ideas here suggest a path to circumvent this.  One could use extremely large qubit-cavity detunings, resulting in both very weak induced cavity nonlinearities, but also weak induced dissipation.  Our scheme shows that such weak nonlinearities could still be harnessed to produce non-classical states.

In this work, we have described a new basic route to generating photonic states that are blockaded:  they have a sharp cutoff in their photon-number distribution, having zero probability to have more than $r$ photons in the state.  This is accomplished by using standard tools (a weak Kerr nonlinearity, one and two photon drives) to realize an effective non-linear drive, cf.~Eq.~(\ref{eq:H-blockade-target}).    In stark contrast to the well-studied unconventional photon blockade mechanism \cite{liew_single_2010}, our scheme can generate truly blockaded states, and states that do not need to be infinitely close to being vacuum. In principle, our basic mechanism is effective even for arbitrarily weak nonlinearities $U \ll \kappa$.  In practice, limitations will arise from the inability to perfectly match the one and two photon drive amplitudes, and the inability to apply the required displacement transformations perfectly.  We showed that the scheme nonetheless can be effective even if these imperfections are present.  

While our analysis focused on generating states that approximate single-photon Fock states, the idea is much more general.  By picking the parameter $r$ in Eq.~(\ref{eq:H-blockade-target}) to be an integer larger than one (which then influences the choice of drive amplitudes via Eqs.~(\ref{eq:Lam1-parameter}) and (\ref{eq:Lam2-parameter})), one can generate higher-order blockaded states:  states that are confined to the manifold spanned by Fock states $|0\rangle, |1 \rangle, ..., |r \rangle$.  Further, the same basic idea can used to generate non-classical, multi-mode entangled states.  One again realizes the nonlinear driving Hamiltonian in Eq.(\ref{eq:H-blockade-target}) in a displaced frame, but now the mode $\hat{a}$ is actually a collective mode of 2 or more distinct cavity modes.  Generating a Fock state in this collective mode directly corresponds to a $W$-style entangled state.  More details are provided in App.~\ref{app:many-body-blockade}.

In summary, we believe that the mechanism discussed here will prove to be valuable tool for generating non-classical photonic states in a variety of platforms where only weak nonlinearities are achievable.  It could also conceivably be harnessed as a tool for quantum simulation, i.e.~to realize models of strongly interacting photons.  Our ideas are compatible with a wide variety of bosonic systems, including optical and microwave cavities, as well as more general superconducting circuit QED setups.

\section{Materials and Methods}

\noindent {\bf RWA Hamiltonian.}
A crucial result of this work is that to implement the nonlinear photon drive of Eq.~(\ref{eq:H-blockade}), we require only a single mode of a bosonic resonator with a weak self-Kerr nonlinearity $U$ and standard one and two photon drives. The starting lab-frame Hamiltonian is thus ($\hbar=1$)
\begin{align}
    \hat{H} &= \omega_c \hat{a}^\dagger \hat{a} + \frac{U}{6}(\hat{a} + \hat{a}^\dagger)^4 \nonumber \\
    &+ (\Lambda_{1} e^{-i\omega_1 t} + \Lambda_{1}^* e^{i\omega_1 t})(\hat{a} + \hat{a}^\dagger) \nonumber \\
    &+ (\Lambda_{2} e^{-i \omega_2 t} + \Lambda_{2}^* e^{i \omega_2 t})(\hat{a}\hat{a} + \hat{a}^\dagger\hat{a}^\dagger).
    \label{eq:lab-frame-H}
\end{align}
Note that the only nonlinearity in this Hamiltonian is Kerr interaction $U$, which we will allow to be extremely weak, i.e. $U \ll \kappa$, where $\kappa$ is the cavity loss rate. The two photon drive $\Lambda_2$ is a standard parametric drive, and can be realized {\it without} requiring a strong single-photon nonlinarity.

We choose the drive frequencies to satisfy  $\omega_2 = 2\omega_1 = 2 (\omega_c - \Delta)$, implying they are equally detuned from the resonance by an amount $\Delta$.
We also work in the standard regime where $\omega_c$ is the largest frequency in the problem, allowing us to make a rotating wave approximation (RWA) on both the nonlinearity and drive terms.  Making the RWA and working in the rotating frame set by the drive frequency  $\omega_1$, we obtain Eq.~(\ref{eq:rwa-H}). Note that we have normal-ordered the nonlinearity; thus the nonlinearity strength in $\hat{H}_{\rm RWA}$ is $U$. Note also that normal-ordering shifts the resonance to $\tilde{\omega}_c = \omega_c + 2 U$; we implicitly assume the detuning from resonance in $\hat{H}_{\rm RWA}$ is thus $\Delta = \omega_1 - \tilde{\omega}_c$.

\noindent {\bf Displacement transformation.}
We use strong driving to enhance the effects of $U$ in $\hat{H}_{\rm RWA}$. We also include single photon loss at a rate $\kappa$ using a standard Lindblad master equation description.  Letting $\hat{\rho}$ denote the reduced density matrix of the cavity mode, we have  
\begin{equation}
    \frac{d}{dt}\hat{\rho} = -i[\hat{H}_{\rm RWA},\hat{\rho}] + \kappa\mathcal{D}[\hat{a}]\hat{\rho}
    \label{eq:blockade-qme}
\end{equation}
where $\mathcal{D}[\hat{a}]\hat{O} = (\hat{a}\hat{O}\hat{a}^\dagger - \{\hat{a}^\dagger\hat{a},\hat{O}\}/2)$ is the standard Lindblad dissipative superoperator. 

The trick is now to show that with appropriate parameter tuning, a simple displacement of our weakly non-linear Hamiltonian in Eq.~(\ref{eq:rwa-H}) can yield exactly the kind of nonlinear driving interaction we are looking for.  In particular, we want a Hamiltonian that is {\it unitarily equivalent} to $\hat{H}_{\rm target}$
in Eq.~(\ref{eq:H-blockade-target}),
where the parameter $r$ will be set to a positive integer.  This Hamiltonian describes a nonlinear driving process that can pump up an initial vacuum state to the $n=r$ Fock state, but no higher. 

To achieve this equivalence, we consider a displacement transformation to a new frame where the original photonic vacuum is shifted to the coherent state $|-\alpha \rangle$; we leave the amplitude $\alpha$ unspecified for the moment.  
This required unitary is 
$\mathcal{D}_\alpha = \exp{(\alpha\hat{a}^\dagger - \alpha^*\hat{a})}$, which  transforms the lowering operator as $\hat{a}\to\hat{a}+\alpha$.  
In this new displaced frame, the master equation for our system has the same form as Eq.~(\ref{eq:blockade-qme}), but with a modified displaced Hamiltonian $\hat{H}_{\alpha}$
\begin{align}
    \hat{H}_\alpha &= U\hat{a}^\dagger\hat{a}^\dagger\hat{a}\hat{a} + \tilde{\Delta} \hat{a}^\dagger\hat{a} \nonumber\\
    &+ (\tilde{\Lambda}_1\hat{a}^\dagger + \tilde{\Lambda}_2\hat{a}^\dagger\hat{a}^\dagger + \tilde{\Lambda}_3\hat{a}^\dagger\hat{a}^\dagger\hat{a} + {\rm h.c.}).
    \label{eq:displaced-H}
\end{align}
All of the terms in the original lab-frame Hamiltonian appear in $\hat{H}_\alpha$, but with altered coefficients; we also generate the desired nonlinear single-photon driving term $\tilde{\Lambda}_3$.  The displaced-frame Hamiltonian parameters are:
\begin{subequations}
    \begin{align}
        \tilde{\Delta} &= \Delta + 4U|\alpha|^2, \\
        \tilde{\Lambda}_1 &= \Lambda_{1} + \alpha\Delta + 2\alpha^*\Lambda_{2} + 2U|\alpha|^2\alpha
    - \frac{1}{2}i\kappa\alpha, \\
        \tilde{\Lambda}_2 &= \Lambda_{2} + U\alpha^2,\\
        \tilde{\Lambda}_3 &= 2U\alpha.
        \label{eq:Lam3Tuning}
    \end{align}
\end{subequations}
Notice that by picking the displacement $\alpha$ and the lab-frame Hamiltonian parameters $\Lambda_1$, $\Lambda_2$, $\Delta$, we have complete control over all of the displaced-frame Hamiltonian parameters. In particular the choices in Eqs.~(\ref{eq:alpha-parameter}-\ref{eqs:OptimalParams}) lead to the target Hamiltonian $\hat{H}_{\rm target}$.
Also notice that the displacement transformation modifies the Lindblad dissipator as $\mathcal{D}[\hat{a}]\mapsto\mathcal{D}[\hat{a}+\alpha]$ which we rewrite as $\mathcal{D}[\hat{a}] - i[(-i\alpha\kappa/2)\hat{a}^{\dagger} + {\rm h.c.},\rho]$. The induced coherent linear drive component has been absorbed into $\tilde{\Lambda}_1$ in Eq. (13b).  The net result is that the damping rate of the cavity is the same in the displaced frame.


\noindent {\bf Photon blockade in the infinite-time steady state.}
The main focus of our work is understanding Fock state generation using the dynamics of Eq.~(\ref{eq:blockade-qme}) for times much shorter than the full relaxation time of the system.  Here, we comment on features of the infinite-time steady state.  The properties of this steady state were discussed in Ref.~\cite{roberts_driven-dissipative_2020} using an exact-solution technique.

\begin{figure}[ht]
    \centering
    \includegraphics[width=0.999\columnwidth]{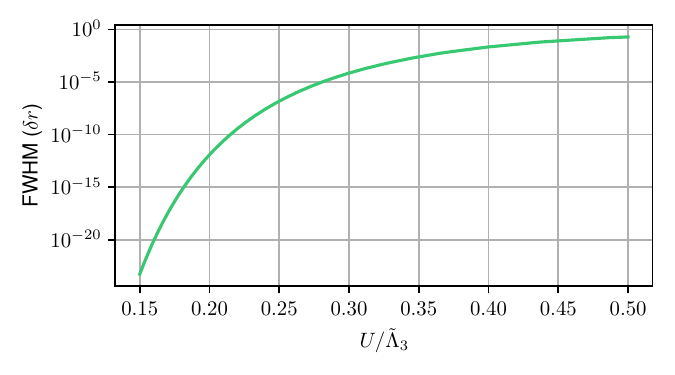}
    \caption{{\bf Steady state photon number antiresonance width as a function of $U/\tilde{\Lambda}_3$}. Full width at half maximum of the steady state photon number at the single-photon blockade antiresonance as a function of $U/\tilde{\Lambda}_3$ as measured by the small deviation $\delta\lambda_1$ (cf. Eq.~\eqref{eq:dLam1Definition}).}
    \label{fig:app-a}
 \end{figure}

As discussed, when $r$ is exactly tuned to an integer, the steady state exhibits blockade:  the steady state photon number distribution truncates at $n=r$.  Surprisingly, this blockade phenomena is lost even for extremely small deviations of $r$ away from an integer.  This manifests itself as an anti-resonance phenomena when the average photon number in the steady state, $\langle \hat{n} \rangle_{\rm ss}$ is plotted versus $r$.  There is a sharp dip in this quantity when $r$ is an integer, with the width of these features $\Delta r \ll 1$.  This behaviour is illustrated in Fig.~\ref{fig:app-a}, where we plot the full-width half-max $\Delta r$ for the anti-resonance in       $\langle \hat{n}(r) \rangle_{\rm ss}$ centered at $r=1$.  We plot this width as a function of  $\tilde{\Lambda}_3/U$.  The plot shows an exponential dependence on this parameter.  Away from the blockade point $r=1$, the steady state photon number is approximately constant and 
has a large value $\gg 1$.

Both the large average steady state photon number away from integer $r$, and the extremely small anti-resonance widths can be understood starting with a semiclassical analysis, which reveals a large-amplitude metastable state.  The semiclassical equation of motion for the amplitude $\alpha = \langle \hat{a}\rangle$ that follows from Eq.~(\ref{eq:blockade-qme}) is:
\begin{equation}
    \frac{d}{dt}\alpha=-2iU\alpha^{*}\alpha^{2}-2i\tilde{\Lambda}_{3}\alpha^{*}\alpha-i\tilde{\Lambda}_{3}\alpha^{2}+i\tilde{\Lambda}_{3}r-\frac{\kappa}{2}\alpha
    \label{eq:semiclassical-eom}
\end{equation}
For $r=0$ (nonlinear drive only) and $\kappa=0$, the steady-state solutions to this equation are $\alpha_0 = 0$ (with multiplicity 2) and
\begin{equation}
    \alpha_{\rm ha} = -\frac{3\tilde{\Lambda}_3}{2U}.
\end{equation}
Because we always assume a regime where $U \ll \tilde{\Lambda}_3$, this amplitude is typically very large.  Including non-zero $\kappa$ and $r$,
we find that the first-order correction to this amplitude is small.  To first order, we have
\begin{equation}
    \alpha_{\rm ha} = -\frac{3\tilde{\Lambda}_{3}}{2U}-\frac{i\kappa}{2\tilde{\Lambda}_{3}}+\frac{2U}{9\tilde{\Lambda}_{3}}r.
    \label{eq:alpha-ha}
\end{equation}
We can also confirm that this is an accurate description of the large-amplitude state by numerically finding the fixed points to Eq.~\eqref{eq:semiclassical-eom} without assuming small $r, \kappa$.

Next, we show that this semiclassical solution is stable by performing a standard linear stability analysis of the semiclassical equations.  
The eigenvalues of the linearized equations of motion for $\alpha$ and $\alpha^*$ about $\alpha_{\rm ha}$ are
\begin{equation}
    \lambda_\pm = -\frac{\kappa}{2}\pm i3\sqrt{3} \frac{\tilde{\Lambda}_3^2}{2U}\left( 1 - \frac{8}{27}\frac{U^2}{\tilde{\Lambda}_3^2}r \right)
    \label{eq:semicl-amp}
\end{equation}
which have negative real parts, indicating linear stability at the semiclassical level. 
Turning to the quantum problem, our system always has a unique steady state, which for integer $r$ is a blockaded state.  Hence, for integer $r$ the above semiclassically-stable state is only unstable due to quantum effects  (i.e.~precisely the blockade physics we have described, which is intimately tied to the discreteness of photon number).

Returning to the quantum problem, we find that upon numerically diagonalizing $\hat{H}_{\rm block}$ in Eq.~\eqref{eq:H-blockade}, there is an eigenstate $|\Phi\rangle$ with photon number $\langle \hat{n} \rangle_\Phi \approx |\alpha_{\rm ha}|^2$ where $\alpha_{\rm ha}$ is given by Eq.~\eqref{eq:semicl-amp}. Focusing on the single photon blockade, $r=1$, we numerically diagonalize the Liouvillian Eq.~\eqref{eq:blockade-qme} which reveals that there is generically a single nonzero eigenvalue $\gamma_{\rm slow}$ which is significantly smaller than $\kappa$; all other decay rates are order $\kappa$ or larger. We seek to show that this eigenvalue corresponds to the decay of the Hamiltonian eigenstate $|\Phi\rangle$ and that the value is exponentially small in $\tilde{\Lambda}_3/U$.

Working under the assumption that $|\Phi\rangle$ is the state whose decay is given by the Liouvillian eigenvalue $\gamma_{\rm slow}$, we use first order degenerate Liouvillian perturbation theory to estimate $\gamma_{\rm slow}$. The exact eigenstates within the single photon blockade manifold $\{|0\rangle,|1\rangle\}$ are given by
\begin{equation}
    |\psi_\pm\rangle = \frac{1}{\sqrt{2}}(|0\rangle \pm |1\rangle);\quad E_\pm = \mp \tilde{\Lambda}_3.
    \label{eq:blockade-eigenstates}
\end{equation}
Note that these span the $\{|0\rangle,|1\rangle\}$ manifold so that $\langle 0|\Phi\rangle = \langle 1|\Phi\rangle=0$. Using the numerically computed $|\Phi\rangle$, we find that it is reasonably well approximated by the coherent state $|\alpha_{\rm ha}\rangle$ with overlap $|\langle \alpha_{\rm ha}|\Phi\rangle|^2 > 0.96$ for $U\ll\tilde{\Lambda}_3$. We enforce orthogonality with the blockade eigenstates Eq.~\eqref{eq:blockade-eigenstates} which gives us the approximate eigenstate 
\begin{equation}
    |\phi\rangle = \frac{1}{\mathcal{N}}\left(|\alpha_{\rm ha}\rangle - e^{-\frac{1}{2}|\alpha_{\rm ha}|^2}|0\rangle - \alpha_{\rm ha}e^{-\frac{1}{2}|\alpha_{\rm ha}|^2}|1\rangle\right)
    \label{eq:approx-ha-eig}
\end{equation}
where $\mathcal{N}$ is the normalization constant. Under the assumption that $|\phi\rangle$ is an approximate eigenstate of $\hat{H}_{\rm block}$, the relevant three-eigenstate degenerate manifold of the unperturbed Liouvillian $\mathcal{L}_0 = -i[\hat{H}_{\rm block},\cdot]$ is $\{|\psi_+\rangle\langle\psi_+|$, $|\psi_-\rangle\langle\psi_-|$, $|\phi\rangle\langle\phi|\}$. (The third \emph{exact} eigenstate is $|\Phi\rangle\langle\Phi|$ of course.) The perturbation is single photon loss
\begin{equation}
    \mathcal{L}_1 = \kappa \mathcal{D}[\hat{a}]
\end{equation}
where $\mathcal{D}[\hat{X}]$ is the standard Lindblad dissipator. We diagonalize the three state subspace with respect to $\mathcal{L}_1$ and compute the eigenvalues. The irrelevant eigenvalues are $\gamma_0 = 0$, whose eigenvector is the $\kappa\ll\tilde{\Lambda}_3$ limit of the single photon blockade steady state, and $\gamma_1 = \kappa/2$, whose eigenvector describes population imbalance relative to the steady state. The final eigenvalue is the only one whose eigenvector involves $|\phi\rangle\langle\phi|$ and for $U\ll\tilde{\Lambda}_3$ is given by
\begin{equation}
    \gamma_{\rm slow} \approx \kappa |\alpha_{\rm ha}|^2\left(1+2|\alpha_{\rm ha}|^2\right)e^{-|\alpha_{\rm ha}|^2}.
    \label{eq:slow-rate}
\end{equation}
This shows that the dissipative gap of the blockade Liouvillian spectrum is exponentially small in $U/\tilde{\Lambda}_3 \ll 1$ due to a quasistable eigenstate of the coherent Hamiltonian.  We thus have provided a quantitative and intuitive understanding of the surprising sensitivity of the steady state to small deviations of $r$ away from integer values, explaining the surprisingly sharp anti-resonance phenomena found in Ref.~\cite{roberts_driven-dissipative_2020}.

\noindent {\bf Estimation of $\Gamma_{\rm esc}$.}
We provide details here on how to use Fermi's Golden Rule (FGR) to estimate the slow rate $\Gamma_{\rm esc}$ (c.f.~Eq.~(\ref{eq:escape-rate})) which governs escape from the blockaded subspace in the presence of imperfect drive amplitudes.  Consider first the simple case where 
$\kappa \ll \tilde{\Lambda}_3$.  We write the system Hamiltonian $\hat{H} = 
\hat{H}_0 + (\delta \tilde{\Lambda}_1 \hat{a}^\dagger + {\rm h.c.})$, where $\hat{H}_0$ is the ideal Hamiltonian with perfect drive amplitude matching (i.e.~$\hat{H}_0 = \hat{H}_{\rm target}$ with $r=1$, c.f.~Eq.~(\ref{eq:H-blockade-target})).  Treating the last term as a perturbation, and letting $|\phi_j \rangle$ ($E_j$) denote eigenstates (eigenvalues) of $\hat{H}_0$, application of FGR yields
\begin{equation}
    \Gamma_{\rm esc} = \sum_{j\in\{{\rm unblock}\}} |\langle\phi_j |\delta \tilde{\Lambda}_1\hat{a}^\dagger |\phi_\pm\rangle|^2 \frac{\gamma_j/2}{(\Delta E)^2 + \gamma_j^2/4}.
    \label{eq:escape-rate-small-kappa}
\end{equation}
Here $|\phi_\pm \rangle$ are the two blockade-subspace eigenstates of $\hat{H}_0$, and $\Delta E = E_j - E_\pm$.  The last factor 
in Eq.~(\ref{eq:escape-rate-small-kappa})
corresponds to the lifetime-broadened density of states of each unblockaded eigenstate; for weak $\kappa$, the decay rate $\gamma_j = \kappa \langle \phi_j | \hat{a}^\dagger \hat{a} | \phi_j \rangle$.  This general form matches that of Eq.~(\ref{eq:escape-rate}) in the main text, with a prefactor $c$ that in general depends on the unblockaded eigenstates of $\hat{H}_0$ and hence $U / \tilde{\Lambda}_3$. We find good agreement between  Eq.~(\ref{eq:escape-rate-small-kappa}) (computed from exact diagonalization) and the rate extracted from numerical simulations of the system dynamics for weak $\kappa$. As an example, we consider $\tilde{\Lambda}_3 = 100 \kappa$. For $U/\tilde{\Lambda}_3 = 0.2$, the estimate is $c = 0.0051$ and the extracted value from the dynamics is $c = 0.0047$, and for $U/\tilde{\Lambda}_3 = 0.3$ the estimate is $c = 0.0036$ and the extracted value is $c = 0.0045$. These are typical of this parameter regime.  The small value of $c$ here directly reflects the delocalization of the unblockaded eigenstates.

For more general regimes, it is trickier to directly apply FGR, as one can no longer treat the effects of $\kappa$ by simply lifetime broadening each unperturbed eigenstate.
For 
$\kappa \gg \tilde{\Lambda}_3$, on can use the fact that the large dissipation will disrupt the formation of coherent eigenstates outside the blockaded subspace.  In this case, we can 
estimate $\Gamma_{\rm esc}$ by considering a transition from either $|\phi_{\pm} \rangle$ to the Fock state $|2\rangle$, whose decay rate is simply $2 \kappa$.  This leads to an approximate decay rate corresponding to Eq.~(\ref{eq:escape-rate}) with parameter-independent constant $c=1$.
For the most relevant regime 
$\kappa \sim \tilde{\Lambda}_3$, it is difficult to rigorously calculate the decay rate as neither $\kappa$ nor the unblockaded coherent dynamics can be treated perturbatively.  As discussed in the main text, numerically a good agreement is found to the general form in Eq.~(\ref{eq:escape-rate}) with $c \sim 0.25$.  Heuristically, this is consistent with the results presented above; the slightly smaller value of $c$ corresponds to the partial delocalization of unblockaded eigenstates.

\bibliographystyle{apsrev4-1}
\bibliography{Lam3-photon-blockade}

\section*{Acknowledgements}

We thank Hong Tang of Yale University for useful conversations. This work was primarily supported by the University of Chicago Materials Research Science and Engineering Center, which is funded by the National Science Foundation under Grant No. DMR-1420709.  This work was also supported by the Air Force Office of Scientific Research MURI program, under Grant No. FA9550-19-1-0399, 
and by the Army Research Office under grant W911-NF-19-1-0380.  It was completed in part with resources provided by the University of Chicago’s Research Computing Center.   AC also acknowledges support from the Simons Foundation through a Simons Investigator award.

\newpage


\appendix


\section{Displacement error modeling and discussion of phase noise}
\label{app:error-modeling}

The protocol for generating single photons (see Fig.~5 in the main text) requires a rapid displacement of the cavity by an amplitude $\alpha_b$ (see Eq.~(4) in the main text) at the beginning and a rapid displacement  by $-\alpha_b$ at the end of the protocol. These displacements will not be perfect in any experiment so we must model small errors in them. Instead of enumerating specific possible errors that can occur during the displacements, we elected to offer a general error model via the Gaussian additive thermal noise channel.
The thermal noise model not only provides results on how the correlation function $g^{(2)}(0)$ is bounded by the initial thermal occupation of the cavity, it can also be directly connected to errors in the displacement parameter $\alpha_b$.  As we show, this model also allows one to directly assess the impact of classical displacement noise.  We also discuss how an almost identical approach can be used to model classical phase noise.

Displacement errors occur when the displacement amplitude is not $\alpha_b$ but some 
\begin{equation}
    \tilde{\alpha}_b = \alpha_b + \delta
\end{equation}
for some small complex error $\delta$. We treat these errors generally by letting $\delta$ be a complex zero-mean Gaussian random variable with variance $\sigma^2$. Averaging over these random displacements starting from vacuum $|0\rangle$ results in a thermal state [30]; or in our case, a thermal state displaced by $\alpha_b$. The occupation number $\bar{n}_{\rm th}$ of the thermal state is related to the variance $\sigma^2$ of the random displacement errors by
\begin{equation}
    \bar{n}_{\rm th} = \sigma^2.
    \label{eq:add-displ-error-equiv}
\end{equation}
One can see this by comparing the variance of the cavity quadratures for the thermal state and for the Gaussian average of the displaced coherent states. To model the displacement errors during the initial displacement numerically, we perfectly displace a thermal state with some occupation $\bar{n}_{\rm th}$:
\begin{equation}
    |0\rangle\langle 0| \mapsto \hat{D}(\alpha_b)\hat{\rho}_{\rm th}(\bar{n}_{\rm th})\hat{D}^\dagger(\alpha_b)
\end{equation}
where $\hat{D}$ is the displacement operator.

At the end of the protocol we must displace the cavity back to the origin and apply the thermal noise channel.
Applying the thermal noise channel to the blockaded state is not so easy because it is no longer a coherent state. Fortunately we are not particularly interested in the exact form of the final state, we are only interested in the correlation functions $\langle \hat{a}^\dagger \hat{a} \rangle$ and $\langle \hat{a}^\dagger \hat{a}^\dagger \hat{a} \hat{a} \rangle$ as these are all we need to compute $g^{(2)}(0)$. Thus at the end of the protocol, we perform a perfect displacement $-\alpha_b$, then we inject the same thermal noise into the final state by passing the cavity mode $\hat{a}$ through a beamsplitter with transmissivity $1-\epsilon$. A mode $\hat{b}$ with a thermal state, occupation $\langle \hat{b}^\dagger\ \hat{b}\rangle = \tilde{n}$, is put on the vacuum port of the beamsplitter. The output mode is thus
\begin{equation}
    \hat{c} = \sqrt{1-\epsilon}\hat{a} + \sqrt{\epsilon}\hat{b}.
\end{equation}
We will consider the limit $\epsilon \to 0$, $\tilde{n}\to\infty$ such that $\epsilon\tilde{n} = \bar{n}_{\rm th}$ is fixed. Now assuming $\langle \hat{a}\rangle = 0$, a reasonable assumption as the coherence between $|0\rangle$ and $|1\rangle$ decays very quickly during the protocol, we find
\begin{equation}
    \langle \hat{c}^\dagger\hat{c} \rangle = \langle \hat{a}^\dagger\hat{a} \rangle + \bar{n}_{\rm th},
\end{equation}
\begin{equation}
    \langle \hat{c}^\dagger\hat{c}^\dagger\hat{c}\hat{c} \rangle = \langle \hat{a}^\dagger\hat{a}^\dagger\hat{a}\hat{a} \rangle + 4 \langle\hat{a}^\dagger\hat{a}\rangle \bar{n}_{\rm th} + 2\bar{n}_{\rm th}^2,
\end{equation}
after taking the limit $\epsilon \to 0$, $\tilde{n}\to\infty$. Finally for small $\bar{n}_{\rm th}\ll \langle\hat{a}^\dagger\hat{a}\rangle$, we compute $g^{(2)}(0) = \langle \hat{c}^\dagger\hat{c}^\dagger\hat{c}\hat{c} \rangle/\langle\hat{c}^\dagger\hat{c}\rangle^2$ to place the bound
\begin{equation}
    g^{(2)}(0) \geq \frac{\langle \hat{a}^\dagger\hat{a}^\dagger\hat{a}\hat{a} \rangle}{\langle\hat{a}^\dagger\hat{a}\rangle^2} + \frac{4 \bar{n}_{\rm th}}{\langle\hat{a}^\dagger\hat{a}\rangle}.
\end{equation}
In the numerical simulations, we perform a perfect displacement back to the origin at the end of the protocol and use this expression to bound $g^{(2)}(0)$.

The Gaussian thermal noise model as presented above describes the additive noise of small displacement errors $\tilde{\alpha}_b  = \alpha_b + \delta$. The results presented in Fig.~1 of the main text are thus readily interpreted as additive displacement errors via Eq.~\eqref{eq:add-displ-error-equiv}. Already this allows an experimentalist to determine the maximum allowed fractional error of the displacements to achieve a desired $g^{(2)}(0)$. Of course, with all other parameters fixed, this maximum fractional error is dependent on the displacement $\alpha_b$. 

In some experimental settings, a more natural error model would correspond to the noise-corrupted displacement having the form
\begin{equation}
    \tilde{\alpha}_b = (1+\delta)\alpha_b.
\end{equation}
This could arise, e.g., because the coherent tone source, e.g., optical laser or microwave generator, will have a limited precision to which its output power and phase can be controlled. Again if $\delta$ is a complex Gaussian random variable, multiplicative displacement errors correspond to the Gaussian thermal channel via
\begin{equation}
    \bar{n}_{\rm th} = |\alpha_b|^2\sigma^2
\end{equation}
where $\sigma^2$ is the variance of $\delta$. Thus the standard deviation $\sigma$ is the fractional error $|\tilde{\alpha}_b - \alpha_b|/|\alpha_b|$ of the displacement operation.

 \begin{figure}[ht]
    \centering
    \includegraphics[width=0.999\columnwidth]{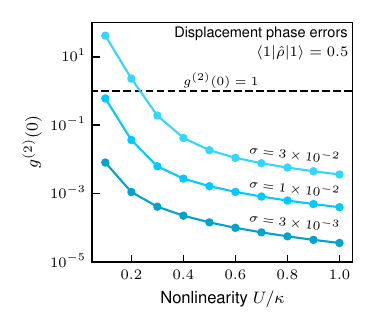}
    \caption{Numerical simulations of the performance of the single photon generation protocol in the presence of displacement phase errors during the initial/final displacement operators. Parameters are chosen such that the effective nonlinear drive amplitude $\tilde{\Lambda}_3 = 2\kappa$ and the final state has $\langle 1|\hat{\rho}|1\rangle = 0.5$. The parameter $\sigma$ is the standard deviation of the phase noise in radians. Note that $g^{(2)}(0)$ must be greater than $(U/\kappa)^2\sigma^2$.}
    \label{fig:app-b}
 \end{figure} 

One drawback of the thermal noise model treats all displacement errors equally: it assumes that amplitude errors $|\tilde{\alpha}_b| = |\alpha_b|(1 + {\rm Re}\,\delta)$ and phase errors $\tilde{\alpha}_b = e^{i{\rm Im}\,\delta}\alpha_b\approx|\alpha_b|(1+i{\rm Im}\,\delta)$ have the same variance. In many settings however, the dominant error will be a phase error.  Thus in what follows we also model the impact of phase errors that are Gaussian distributed. Without loss of generality we let $\alpha_b = |\alpha_b|$. The phase error is
\begin{equation}
    \tilde{\alpha}_b = e^{i\theta}\alpha_b \approx (1 + i\theta)\alpha_b
\end{equation}
where $\theta$ is a real Gaussian random variable with variance $\sigma^2 \ll 1$. We define the cavity quadratures
\begin{align}
    \hat{X} &= \frac{1}{\sqrt{2}}(\hat{a} +\hat{a}^\dagger) \\
    \hat{Y} &= \frac{1}{i\sqrt{2}}(\hat{a} -\hat{a}^\dagger).
\end{align}
To linear order in $\theta$, phase noise causes diffusion in the $\hat{Y}$ quadrature. The variance of this diffusion is $|\alpha_b|^2\sigma^2$ Thus our Gaussian phase noise model diffuses the $\hat{Y}$ quadrature such that its variance is $\langle\Delta\hat{Y}^2\rangle = \frac{1}{2} + |\alpha_b|^2\sigma^2$ while leaving the $\hat{X}$ quadrature variance $\langle\Delta\hat{X}^2\rangle = \frac{1}{2}$. This noise channel is easily applied to the cavity vacuum: squeeze a thermal state along the $\hat{X}$ quadrature such that the $\hat{X}$ quadrature variance is reduced to $\langle\Delta\hat{X}^2\rangle = \frac{1}{2}$ and the $\hat{Y}$ quadrature variance is increased to $\langle\Delta\hat{Y}^2\rangle = \frac{1}{2} + |\alpha_b|^2\sigma^2$, then perfectly displace this state by $\alpha_b$. The map is
\begin{equation}
    |0\rangle\langle 0| \mapsto \hat{D}(\alpha_b)\hat{S}(\xi)\hat{\rho}_{\rm th}(\bar{n}_{\rm th})\hat{S}^\dagger(\xi)\hat{D}^\dagger(\alpha_b)
\end{equation}
where $\hat{S}$ is the squeeze operator. The thermal occupation and squeezing are set by
\begin{align}
    \bar{n}_{\rm th} &= \frac{1}{2}\left(\sqrt{1+2|\alpha_b|^2\sigma^2} -1\right) \\
    \xi &= \frac{1}{4}\ln \left(1+2|\alpha_b|^2\sigma^2\right).
\end{align}

As was the case with the thermal noise model, applying the channel to vacuum is easy but applying it to the blockaded state is not so easy. We thus take the same approach which is to bound the second order coherence. This time, after perfectly displacing the cavity back to the origin, the output mode $\hat{c}$ is given by 
\begin{equation}
    \hat{c} = \hat{a} + \frac{i}{\sqrt{2}}dY
\end{equation}
where $dY$ is a real Gaussian random variable with variance $|\alpha_b|^2 \sigma^2$. This describes diffusion in the $\hat{Y}$ quadrature of the output mode $\hat{c}$. Assuming $\langle\hat{a}\rangle = 0$ and $\langle\hat{a}\hat{a}\rangle = 0$, the second order coherence is bounded by
\begin{equation}
    g^{(2)}(0) \geq \frac{\langle \hat{a}^\dagger\hat{a}^\dagger\hat{a}\hat{a} \rangle}{\langle\hat{a}^\dagger\hat{a}\rangle^2} + \frac{2|\alpha_b|^2\sigma^2}{\langle\hat{a}^\dagger\hat{a}\rangle}.
\end{equation}
We compute the correlation functions in $\hat{a}$ at the end of the numerical simulation and use this expression to put a lower bound on $g^{(2)}(0)$.

Shown in Fig.~\ref{fig:app-b} are numerical simulations of the time-dependent single photon generation protocol for various choices of $U/\kappa$ subject to phase noise during the initial and final displacements as described above. The simulation parameters are chosen to produce (in the ideal case) a final blockaded state with $\langle1|\hat{\rho}|1\rangle = 0.5$. The curves are labeled by the standard deviation $\sigma$ of the phase noise in radians.

\section{Multi-mode generalization and preparation of non-Gaussian entangled states}
\label{app:many-body-blockade}

Our basic scheme in Eq.~\eqref{eq:H-blockade} can be easily extended to a mechanism that allows the generation of entangled $M$-mode non-Gaussian photonic states {\it using only weak nonlinearities} $U \ll \kappa$.  For concreteness, we describe here the extension to a $M=2$ mode system with lowering operators $\hat{a}_1, \hat{a}_2$.  The idea is simple:  we again want to realize the nonlinear driving Hamiltonian in Eq~(\ref{eq:H-blockade}) in a displaced frame, but now the single-mode lowering operator $\hat{a}$ by a collective mode, e.g. 
\begin{equation}
    \hat{b} = (\hat{a}_1 + \hat{a}_2)/\sqrt{2}.
    \label{eq:bmode}
\end{equation}
The desired nonlinear driving Hamiltonian is
\begin{equation}
    \hat{H}_{{\rm block},2} = \tilde{\Lambda}\hat{b}^\dagger(\hat{b}^\dagger\hat{b} - r) + {\rm h.c.}. \label{eq:HBlockadeTwoMode}
\end{equation}
In what follows, we focus on realizing a single-excitation state, and hence take $r=1$.  Following the logic of ``Dynamics for ideal drive amplitude matching,'' the ideal dynamics under this Hamiltonian can prepare a single excitation in the collective $b$ mode.  If we use the photon number of each $a_1, a_2$ mode to separately encode a qubit, than the single excitation state produced is a maximally entangled Bell state $\left( |10\rangle + |01\rangle \right) / \sqrt{2}$.

We stress that the Hamiltonian in Eq.~(\ref{eq:HBlockadeTwoMode}) amounts to simply replacing $\hat{a}$ in Eq.~(\ref{eq:H-blockade}) of the main text with the collective mode $\hat{b}$.  It thus immediately follows that if we take the four-wave mixing Hamiltonian Eq.~(\ref{eq:rwa-H}) in the main text, replace $\hat{a}$ with $\hat{b}$, then the resulting Hamiltonian is equivalent to Eq.~(\ref{eq:HBlockadeTwoMode}) up to unitary displacement operators of modes $\hat{a}_1$ and $\hat{a}_2$.

Given this, a simple substitution then in Eq.~(\ref{eq:rwa-H}) yields the required form of the starting two-mode Hamiltonian
\begin{align}
    \hat{H}_{{\rm RWA},2} = U\hat{b}^\dagger\hat{b}^\dagger\hat{b}\hat{b} + \Delta \hat{b}^\dagger\hat{b} + (\Lambda_{1}\hat{b}^\dagger + \Lambda_{2}\hat{b}^\dagger\hat{b}^\dagger + {\rm h.c.}).
\end{align}
where again $\hat{b}$ is given by Eq.~(\ref{eq:bmode}).  The linear and quadratic terms that are generated describe linear drives on both modes, detuning and beam-splitter couplings, and parametric drives (both degenerate and non-degenerate).  The nonlinear four-wave mixing terms take the form
\begin{align}
    \hat{H}_{{\rm RWA},2,U} & = 
    \frac{U}{2} \left(
        \sum_{j,k=1}^2 \hat{a}_j^\dagger\hat{a}_k^\dagger\hat{a}_j\hat{a}_k
        + \left( \hat{a}_1^\dagger\hat{a}_1^\dagger\hat{a}_2\hat{a}_2
        + {\rm h.c.} \right)
        \right)
\end{align}
Note that we now require both self and cross Kerr interactions, but also four-wave mixing processes that exchange interactions between modes $1$ and $2$.  

Following the logic of the main text, we imagine displacing each mode $\hat{a}_j$ by the amplitude $\alpha_b$ given in Eq.~(\ref{eq:alpha-parameter}).  
By further tuning the parameters the drive parameters $\Lambda_1, \Lambda_2$ and detuning parameter $\Delta$ as per Eqs.~(\ref{eqs:OptimalParams}), the above two-mode Hamiltonian is unitarily equivalent to the desired two-mode blockade Hamiltonian in Eq.~(\ref{eq:HBlockadeTwoMode}).
We thus show how our basic ideas generalize directly to multi-mode systems; other related approaches are also possible.  While the introduction of modes does involve more complexities, the basic feature of our original scheme remains:  generation of non-classical blockaded (now entangled) states is possible even if the four-wave mixing nonlinearities $U$ are much weaker that photonic loss.

\end{document}